\documentclass[apj,usenatbib]{emulateapj}

\usepackage{amsmath}
\usepackage{graphicx}
\usepackage{hyperref}
\usepackage{gensymb}
\usepackage{textcomp}

\newcommand{\art}{\textit{Simetra}}
\newcommand{\diff}{\, \mathrm{d}}
\newcommand{\psignal}{p(\mathbf{x}\, |\, H_{1})}
\newcommand{\pnoise}{p(\mathbf{x}\, |\, H_{0})}
\newcommand{\amax}{A_{1}}

\newcommand{\rhonorm}{\tilde{\rho}}
\newcommand{\rhostar}{\tilde{\rho}^{\ast}}
\newcommand{\pfa}{P_{FA}}
\newcommand{\imnoise}{\sigma_{\mathrm{im}}}
\newcommand{\aegean}{\textsc{aegean}}
\newcommand{\python}{\texttt{python}}
\newcommand{\wsclean}{\textsc{wsclean}}
\newcommand{\cotter}{Cotter}
\newcommand{\aoflagger}{AOFlagger}
\newcommand{\uvfits}{\textsc{uvfits}}
\newcommand{\thermal}{\mathrm{im}}
\newcommand{\fits}{\textsc{fits}}
\newcommand{\scipy}{\texttt{scipy}}

\shorttitle{Matched Filter and Radio Transient Detection}

\begin{document}

\title{A Matched Filter Technique for Slow Radio Transient Detection and First Demonstration with the Murchison Widefield Array}

\author{%
L.~Feng$^{1}$,
R.~Vaulin$^{1,2}$,
J.~N.~Hewitt$^{1}$,
R.~Remillard$^{1}$,
D.~L.~Kaplan$^{3}$,
Tara~Murphy$^{4,5}$,
N.~Kudryavtseva$^{6}$,
P.~Hancock$^{7,5}$,
G.~Bernardi$^{8}$,
J.~D.~Bowman$^{9}$,
F.~Briggs$^{10}$,
R.~J.~Cappallo$^{11}$,
A.~A.~Deshpande$^{12}$,
B.~M.~Gaensler$^{4,5,13}$,
L.~J.~Greenhill$^{14}$,
B.~J.~Hazelton$^{15}$,
M.~Johnston-Hollitt$^{16}$,
C.~J.~Lonsdale$^{11}$,
S.~R.~McWhirter$^{11}$,
D.~A.~Mitchell$^{17,5}$,
M.~F.~Morales$^{15}$,
E.~Morgan$^{1}$,
D.~Oberoi$^{18}$,
S.~M.~Ord$^{17,5}$,
T.~Prabu$^{12}$,
N.~Udaya~Shankar$^{12}$,
K.~S.~Srivani$^{12}$,
R.~Subrahmanyan$^{12,5}$,
S.~J.~Tingay$^{7,5,19}$,
R.~B.~Wayth$^{7,5}$,
R.~L.~Webster$^{20,5}$,
A.~Williams$^{7}$
and
C.~L.~Williams$^{1}$
\\
$^{1}$~MIT Kavli Institute for Astrophysics and Space  Research, Massachusetts Institute of Technology, Cambridge, MA  02139, USA\\
$^{2}$~Sqrrl Data Inc., 125 CambridgePark Drive, Suite 401, Cambridge, MA 02140, USA\\
$^{3}$~Department of Physics, University of Wisconsin--Milwaukee, Milwaukee, WI 53201, USA\\
$^{4}$~Sydney Institute for Astronomy, School of Physics, The University of Sydney, NSW 2006, Australia\\
$^{5}$~ARC Centre of Excellence for All-sky  Astrophysics (CAASTRO)\\
$^{6}$~Institute of Cybernetics at Tallinn University of Technology, Akadeemia tee 21, 12618 Tallinn, Estonia\\
$^{7}$~International Centre for Radio Astronomy Research, Curtin University, Bentley, WA 6102, Australia\\
$^{8}$~Department of Physics and Electronics, Rhodes University, PO Box 94, Grahamstown 6140, South Africa\\
$^{9}$~School of Earth and Space Exploration, Arizona  State University, Tempe, AZ 85287, USA\\
$^{10}$~Research School of Astronomy and Astrophysics, Australian National University, Canberra, ACT 2611, Australia\\
$^{11}$~MIT Haystack Observatory, Westford, MA 01886, USA\\
$^{12}$~Raman Research Institute, Bangalore 560080, India\\
$^{13}$~Dunlap Institute for Astronomy and Astrophysics, University of Toronto, 50 St.\ George Street, Toronto, ON M5S 3H4, Canada\\
$^{14}$~Harvard-Smithsonian Center for Astrophysics, Cambridge, MA 02138, USA\\
$^{15}$~Department of Physics, University of Washington, Seattle, WA 98195, USA\\
$^{16}$~School of Chemical \& Physical Sciences,  Victoria University of Wellington, Wellington 6140, New Zealand\\
$^{17}$~CSIRO Astronomy and Space Science (CASS), PO Box  76, Epping, NSW 1710, Australia\\
$^{18}$~National Centre for Radio Astrophysics, Tata Institute for Fundamental Research, Pune 411007, India\\
$^{19}$~Osservatorio di Radio Astronomia, Istituto Nazionale di Astrofisica, Bologna, Italy, 40123\\
$^{20}$~School of Physics, The University of Melbourne, Parkville, VIC 3010, Australia}

\begin{abstract}
Many astronomical sources produce transient phenomena at radio frequencies, but the transient sky at low frequencies ($<300$\,MHz) remains relatively unexplored. Blind surveys with new widefield radio instruments are setting increasingly stringent limits on the transient surface density on various timescales. Although many of these instruments are limited by classical confusion noise from an ensemble of faint, unresolved sources, one can in principle detect transients below the classical confusion limit to the extent that the classical confusion noise is independent of time. We develop a technique for detecting radio transients that is based on temporal matched filters applied directly to time series of images rather than relying on source-finding algorithms applied to individual images. This technique has well-defined statistical properties and is applicable to variable and transient searches for both confusion-limited and non-confusion-limited instruments. Using the Murchison Widefield Array as an example, we demonstrate that the technique works well on real data despite the presence of classical confusion noise, sidelobe confusion noise, and other systematic errors. We searched for transients lasting between 2 minutes and 3 months. We found no transients and set improved upper limits on the transient surface density at 182\,MHz for flux densities between $\sim 20$--$200$\,mJy, providing the best limits to date for hour- and month-long transients. 
\end{abstract}



\section{Introduction} \label{introduction}

Many astrophysical systems and physical processes give rise to radio variability and transient phenomena. These range from propagation effects, e.g. extreme scattering events seen in active galactic nuclei \citep{fiedler1987}, to magnetic activity in ultracool dwarfs \citep{hallinan2007} and magnetars \citep{gaensler2005}, to explosive events such as supernovae \citep{soderberg2010} and gamma-ray bursts \citep{frail1997}. In addition, there are transient sources of the unknown nature, such as the Galactic Center radio transients \citep{hyman2005}, fast radio bursts \citep{thornton2013}, and sources with no known counterparts at other wavelengths \citep{thyagarajan2011}; as well as hypothesized sources of radio emission, such as exoplanets \citep{lazio2004}, orphan afterglows \citep{levinson2002}, and gravitational wave counterparts from compact binary coalescence \citep{nakar2011}. 

Despite the numerous observed and predicted radio transient sources, few transients have been detected at radio frequencies $<300$\,MHz. However, recent blind surveys with new widefield radio interferometers, such as the Murchison Widefield Array\footnote{http://mwatelescope.org/} (MWA, \citealt{lonsdale2009}, \citealt{tingay2013}) and the Low-Frequency Array\footnote{http://www.lofar.org/} (LOFAR, \citealt{vanhaarlem2013}), are setting increasingly stringent limits on the surface density of radio transients at different sensitivities and over a wide range of timescales. In particular, \citet{bell2014} reported a limit of $<7.5\times 10^{-5}$\,deg$^{-2}$ above 5.5\,Jy at 154\,MHz for minute-to-hour transients; \citet{carbone2014} reported a limit of $<1.0\times 10^{-3}$\,deg$^{-2}$ above 500\,mJy at 152\,MHz for minute-to-hour transients; \citet{cendes2014} reported a limit of $<2.2 \times 10^{-2}$\,deg$^{-2}$ above 500\,mJy at 149\,MHz for 11-min transients; \citet{rowlinson2015} reported a limit of $<6.4 \times 10^{-7}$\,deg$^{-2}$ above 210\,mJy at 182\,MHz for 30-sec transients up to $<6.6 \times 10^{-3}$\,deg$^{-2}$ for 1-yr transients; \citet{polisensky2016} reported a range of limits ($\sim 10^{-4}$\,deg$^{-2}$ above $100$\,mJy) for 10-min and 6-hour transients at 340\,MHz; and \citet{murphy2016} reported a limit of $<1.8 \times 10^{-4}$\,deg$^{-2}$ above 100\,mJy between 150--200\,MHz for transients that lasted 1--3\,yr. \citet{jaeger2012} reported a transient detection that corresponded to a surface density of 0.12\,deg$^{-2}$ above 2.1\,mJy over 12\,hours at 325\,MHz, while \citet{stewart2016} reported a transient detection that implied a surface density of $1.5\times 10^{-5}$\,deg$^{-2}$ above $7.9$\,Jy for $\sim 10$-min transients at 60\,MHz.

These surveys, searching for slow transients that are on the timescale of a snapshot or longer (generally $\gtrsim$~min), are usually limited by the image root-mean-square (RMS) noise and the detection threshold of a source-finding algorithm. One component of the RMS noise is the classical confusion noise $\sigma_{c}$, which arises from a background of faint, unresolved sources \citep{condon1974}. This is a spatial noise in the image that depends on the source distribution in the sky and the instrument resolution:
\begin{equation}\label{eq:confusion_def}
\sigma_{c}^{2} = \Omega_{b} \int_{0}^{S_{c}} S^{2} \frac{\diff n}{\diff S} \diff S,
\end{equation}
where $\Omega_{b}$ is the solid angle of the synthesized beam, $S$ is the source flux density, and $\diff n/\diff S$ is the differential number density of sources \citep{thyagarajan2013}. The upper limit of integration $S_{c}$ is the flux density of a source detected at a particular signal-to-noise ratio $q = S_{c} / \sigma_{c}$; usually $S_{c}$, referred to as the confusion limit, is determined iteratively until $q = 5$. 

As the instrument resolution improves and more sources are resolved, the classical confusion noise decreases. For a modest angular resolution like that of the MWA at $\sim 2'$, however, the classical confusion noise can become a limiting factor in the image RMS noise because the classical confusion noise does not decrease with longer integration times. This, in turn, limits the ability of source-finding algorithms to identify faint sources in an image. The theoretical estimate of the confusion limit for the MWA is $\sim 6$\,mJy at 150\,MHz and $5'$ angular resolution \citep{tingay2013}, while $P(D)$ analysis, using different source count estimates, suggests that the classical confusion noise for the MWA at $154$\,MHz is $\sim 1.7$\,mJy\,beam$^{-1}$ at $2.3'$ angular resolution \citep{franzen2016}. 

The classical confusion noise is largely independent of time, however, so it is, in principle, not a limit for detecting fainter but varying brightness \citep{condon1974}. A simple method for detecting brightness variations below the classical confusion noise is image subtraction. For example, one can subtract images taken at the same local sidereal time to remove both classical confusion noise and sidelobe confusion noise, the latter of which is due to residual synthesized beam sidelobes. For many surveys, however, the images are taken at different local sidereal times, and sidelobe confusion noise can dominate. In this case, image subtraction can be prone to artifacts. Even without image subtraction, CLEAN artifacts impact radio transient searches. For instance, \citet{frail2012} found that many transient candidates reported by \citet{bower2007} were in fact artifacts or the result of calibration errors, and those that were not identified as such were detections at lower significance. A technique that could measure or account for the distribution of artifacts would be able to identify astrophysical transient candidates more reliably. 

As many radio transients are expected to be faint ($\lesssim$\,mJy; see \citealt{metzger2015}), we seek a method that not only takes advantage of the time-independent nature of the classical confusion noise to detect transients without relying on source-finding algorithms that could otherwise limit the sensitivity of the search, but also has well-defined statistical properties that take into account the distribution of artifacts. 

We adapted the well-known matched filter technique \citep{helmstrom1968} to detect radio transients in the presence of classical confusion noise by drawing on the experience of the LIGO community (Laser Interferometer Gravitational-Wave Observatory), which developed techniques to detect gravitational wave signals in the presence of non-Gaussian noise (e.g. \citealt{biswas2012}, \citealt{allen2012}). This technique operates on the image pixel level, has well-defined statistical properties, and is applicable to variable and transient searches for both confusion-limited and non-confusion-limited instruments. We applied this technique to search for slow transients in the MWA data. In Section~\ref{sec:theory}, we describe the mathematical framework for our radio transient detection technique and derive its statistical properties. In Section~\ref{sec:data}, we describe the MWA data reduction procedure. In Section~\ref{sec:performance}, we discuss the performance of this technique on real MWA data to demonstrate its potential for sensitive transient searches. In Section~\ref{sec:analysis}, we describe the transient search analysis using this technique, discuss the results of our search, and set an improved upper limit on the transient surface density at 182\,MHz, the best to date for hour- and month-long transients. In Section~\ref{sec:conclusion}, we conclude that our technique is capable of detecting faint transients and discuss areas of improvement as well as future work. 


\section{Theory}\label{sec:theory}

The presence of classical confusion noise limits the ability of source-finding algorithms to identify sources with flux densities near or below the confusion limit, but the source population that contributes to the classical confusion noise is independent of time unless they are genuine transient or variable events. Thus a transient detection technique that searches for brightness variations on top of a constant signal without relying on a source-finding algorithm is needed to detect transient signals approaching the confusion limit. In this section we describe a technique that identifies transients in individual image pixels despite the classical confusion noise.

Adapted from matched filter techniques, which have been used in engineering applications \citep{helmstrom1968} and gravitational wave astronomy (\citealt{allen2012}, \citealt{biswas2012}), this technique searches for brightness variations on top of a constant signal in individual pixels without using source-finding algorithms. We derive a new transient detection statistic from this technique, discuss its statistical properties, and relate it to the sensitivity of a radio transient search. Although our formalism is derived for transient detection in the image domain, it is similar to the formalism for source detection in the visibility domain \citep{trott2011}. 

To determine whether or not there is a transient signal in a particular image pixel, we compare two hypotheses: the transient is absent (the null hypothesis $H_{0}$), and the transient is present (the alternative hypothesis $H_{1}$). Usually, $H_{0}$ only includes random noise, but in this case, we add a constant background\footnote{Pure random noise is a special case where the constant background is zero.} to $H_{0}$ to represent the time-independent brightness contribution from confusion sources (or other steady sources) because we are only interested in the change in brightness over time: 
\begin{align}
H_{0}&: x_{i} = c + \sigma_{i}, \\
H_{1}&: x_{i} = c + Af_{i} + \sigma_{i}.
\end{align}
For a fixed pixel, $x_{i}$ is the measured brightness in the $i$th snapshot $(i = 1, 2, \ldots, N)$, $c$ is the constant background that follows the distribution characterized by $\sigma_{c}$, $\sigma_{i}$ is the RMS noise (thermal, sidelobe confusion, and other random errors) measured for each snapshot, and $Af_{i}$ is the transient signal, where $A$ is the overall amplitude for a light curve template $\mathbf{f} \equiv \{f_{1}, f_{2}, \ldots, f_{N} \}$. Given the two hypotheses and the data $\mathbf{x} = \{ x_{1}, x_{2}, \ldots, x_{N} \}$, we compute the ratio of the likelihood functions known as the Bayes factor or the likelihood ratio as part of hypothesis testing (\citealt{neyman1933},\citealt{kass1995}): 
\begin{equation}
\Lambda(\mathbf{x}) = \frac{\psignal}{\pnoise} \label{eq:lambda},
\end{equation}
where $p(\mathbf{x}\, | \, H_{i})$ is the probability of observing $\mathbf{x}$ given that $H_{i}$ is true for $i = (0, 1)$.

To derive an analytical result, we assume that the image noise follows a Gaussian distribution with $\mu = 0$ and $\imnoise = \{ \sigma_{1}, \sigma_{2}, \ldots, \sigma_{N} \}$, but we show later that our transient analysis does not rely on this assumption. For the MWA data, this noise is a combination of thermal noise, (residual) sidelobe confusion noise, and other random errors. The likelihood functions are thus the following: 
\begin{align}
\pnoise &= \int \mathcal{N}_{0} \exp \left( -\sum_{i=1}^{N} \frac{b_{i}^{2} (x_{i} - c)^{2}}{2\sigma_{i}^{2}} \right) p(c\, |\, H_{0}) \diff c \label{eq:pnoise}, \\ 
\begin{split}
\psignal &= \int \mathcal{N}_{1} \exp \left( -\sum_{i=1}^{N} \frac{b_{i}^{2} (x_{i} - c - Af_{i})^{2}}{2\sigma_{i}^{2}} \right) \times \\
& p(c\, |\, H_{1}) p(A\, |\, H_{1}) \diff c \diff A \label{eq:psignal}.
\end{split}
\end{align}
$\mathcal{N}_{0}$ and $\mathcal{N}_{1}$ are the normalization factors for a multivariate normal distribution. $b_{i}$ is the value of the $i$th primary beam for the given pixel. $p(c\, |\, H_{0})$, $ p(c\, |\, H_{1})$, and $p(A\, |\, H_{1})$ are the probability distributions of $c$ and $A$ given the respective hypotheses. We assume flat priors; in other words, we assume that $c$ and $A$ are independent and uniformly distributed, and we estimate them by using a least-squares approach \citep{sivia2006}. To do that, we solve Equations~\ref{eq:pnoise} and \ref{eq:psignal} by approximating the integral with the value at its extremum, i.e. $p(\mathbf{x}\,|\,H_{j}) \approx \mbox{const} \times \exp(-\chi_{\mathrm{min}}^{2}/2)$ where $\chi^{2}_{\mathrm{min}}$ is the solution to $\nabla \chi^{2} = 0$. 
Specifically, for Equation~\ref{eq:pnoise}, where
\begin{equation}
\chi_{0}^{2} \equiv \sum_{i=1}^{N} \frac{b_{i}^{2} (x_{i} - c)^{2}}{\sigma_{i}^{2}},
\end{equation}
its extremum is computed by solving
\begin{equation}
\left.\frac{d\chi_{0}^{2}}{dc}\right|_{c = c_{0}} = \sum_{i=1}^{N} \left.\frac{-2b_{i}^{2} (x_{i} - c)}{\sigma_{i}^{2}}\right|_{c = c_{0}} = 0;
\end{equation}
and for Equation~\ref{eq:psignal}, where
\begin{equation}
\chi_{1}^{2} \equiv \sum_{i=1}^{N} \frac{b_{i}^{2} (x_{i} - c - Af_{i})^{2}}{\sigma_{i}^{2}},
\end{equation}
its extrema are computed by solving the two equations
\begin{align}
&\left.\frac{\partial \chi_{1}^{2}}{\partial c}\right|_{c = c_{1}} = \sum_{i=1}^{N} \left.\frac{-2b_{i}^{2} (x_{i} - c - Af_{i})}{\sigma_{i}^{2}}\right|_{c = c_{1}} = 0 \,\,\,\,\,\mbox{and} \\
&\left.\frac{\partial \chi_{1}^{2}}{\partial A}\right|_{A = \amax} = \sum_{i=1}^{N} \left.\frac{-2f_{i}b_{i}^{2} (x_{i} - c - Af_{i})}{\sigma_{i}^{2}}\right|_{A = \amax} = 0.
\end{align}
We note that the choice of using flat priors in this derivation allows us to obtain analytical solutions and simplifies the computation process to demonstrate the technique. The choice of priors will affect the outcome at low signal amplitudes, but accounting for this effect is beyond the scope of this work.

For $H_{0}$, the solution $c = c_{0}$ is an estimate of the constant background level, including the contribution from confusion sources, and it is given by the weighted average of the data: 
\begin{equation}
c_{0} = \langle \mathbf{x} \rangle \equiv \frac{\sum b_{i}^{2} x_{i} / \sigma_{i}^{2}}{\sum b_{i}^{2} / \sigma_{i}^{2}} \label{eq:c0}.
\end{equation}
For $H_{1}$, the solution $c = c_{1}$ is an estimate of the constant background level in the presence of a transient signal; $c_{1}$ becomes $c_{0}$ in the absence of the transient signal. The other solution $A = A_{1}$, which is unitless, is the amplitude of the transient signal given the predefined template $\mathbf{f}$, which has brightness units.
\begin{align}
c_{1} &= \langle \mathbf{x} \rangle - \amax \langle \mathbf{f} \rangle \label{eq:c1}, \\
\amax &= \frac{(\mathbf{x}, \mathbf{f} - \langle \mathbf{f} \rangle)}{(\mathbf{f} - \langle \mathbf{f} \rangle, \mathbf{f} - \langle \mathbf{f} \rangle)} \label{eq:amax}.
\end{align}
The notation $\langle \mathbf{x} \rangle$ represents the weighted average of $\mathbf{x}$ as in Equation~\ref{eq:c0}, and $(\mathbf{x}, \mathbf{f}) \equiv \sum b_{i}^{2} (x_{i}f_{i}) / \sigma_{i}^{2}$ denotes the ``weighted'' inner product between $\mathbf{x}$ and $\mathbf{f}$. Note that an equivalent way of writing $(\mathbf{x}, \mathbf{f} - \langle \mathbf{f} \rangle)$ is $(\mathbf{x} - \langle \mathbf{x} \rangle, \mathbf{f})$, which we interpret as how well the light curve template matches the data after the constant component is subtracted. However, $(\mathbf{f} - \langle \mathbf{f} \rangle)$ is simpler and computationally less intensive to calculate since $\mathbf{f}$ is predefined and identical for all pixels, so we write Equation~\ref{eq:amax} in its given form. 

\begin{figure}
  \centering
  \includegraphics[width=0.5\textwidth]{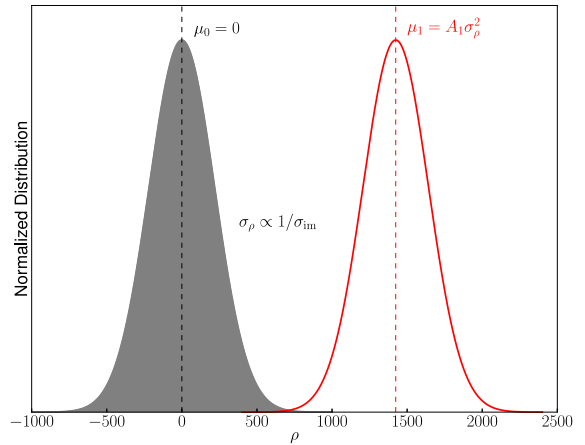}
  \caption[Theoretical distributions of the transient detection statistic]{Theoretical distributions of the transient detection statistic $\rho$: background (gray-filled) and signal (red-line) with arbitrary image noise $\imnoise$ and signal amplitude $A_{1}$. As $\imnoise$ decreases, the widths of the distributions increase according to $\sigma_{\rho} \propto 1/\imnoise$, but the mean of the signal distribution increases faster according to $\mu_{1} \propto 1/\imnoise^{2}$, so the sensitivity improves as there is better separation between background and signal. Similarly, the brighter the signal (larger $A_{1}$), the better the sensitivity.}
  \label{fig:rho_intuition}
\end{figure}

Having determined the best fit values for $c$ and $A$, we substitute Equations~\ref{eq:c0}, \ref{eq:c1}, and \ref{eq:amax} into the approximations of Equations~\ref{eq:pnoise} and \ref{eq:psignal}, which then lets us solve for Equation~\ref{eq:lambda}. As $\Lambda(\mathbf{x})$ is essentially a ratio of exponents normalized by a constant, we rewrite Equation~\ref{eq:lambda} in the form $\Lambda(\mathbf{x}) \approx \mbox{const} \times \exp \left[ \rho^{2}/2\sigma_{\rho}^{2} \right]$ and define the new quantities $\rho$ to be the ``detection statistic'' and $\sigma_{\rho}$ to be the standard deviation of the $\rho$ distribution:
\begin{align}
\rho &= (\mathbf{x}, \mathbf{f} - \langle \mathbf{f} \rangle) = \sum_{i=0}^{N} \frac{b_{i}^{2}}{\sigma_{i}^{2}} x_{i}(f_{i}-\langle \mathbf{f} \rangle)\label{eq:rho}, \\
\sigma_{\rho} &= \sqrt{(\mathbf{f} - \langle \mathbf{f} \rangle, \mathbf{f} - \langle \mathbf{f} \rangle)} = \sqrt{\sum_{i=0}^{N} \frac{b_{i}^{2}}{\sigma_{i}^{2}} (f_{i}-\langle \mathbf{f} \rangle)^{2}} \label{eq:rhonorm}.
\end{align}
$\rho$ is a modified version of the matched filter \citep{helmstrom1968} and determines the likelihood that $\mathbf{x}$ contains the signal $\amax \mathbf{f}$. In other words, it determines which hypothesis is favored (transient absent or present) and by how much. As $\rho$ is a linear superposition of Gaussian random variables, its distribution is also Gaussian with width $\sigma_{\rho}$. When the transient signal is absent, the mean of $\rho$ is $\mu_{0} = 0$. When the transient signal is present, the mean of $\rho$ is shifted by the signal and becomes $\mu_{1} = A_{1} (\mathbf{f} - \langle \mathbf{f} \rangle, \mathbf{f} - \langle \mathbf{f} \rangle) = A_{1} \sigma_{\rho}^{2}$. This is illustrated in Figure~\ref{fig:rho_intuition}.

$\rho$ and $\sigma_{\rho}$ are also related to the more familiar quantity $\imnoise$ through the weighted inner product: $\rho \propto A_{1}/\imnoise^{2}$ and $\sigma_{\rho} \propto 1/\imnoise$. As $\imnoise$ decreases, both $\rho$ (or $\mu_{1}$) and $\sigma_{\rho}$ increase, but $\rho$ (or $\mu_{1}$) increases faster than $\sigma_{\rho}$, which leads to a better separation of background and signal (see Figure~\ref{fig:rho_intuition}) and hence an improved sensitivity. Similarly, the brighter the transient (larger $A_{1}$), the better the sensitivity. 

So far we considered $\rho$ for a single template. However, in a real search, we use a bank of templates and maximize $\rho$ over the various parameters that characterize these templates, such as start times and durations, so the practical statistic is
\begin{equation}
\rhonorm = \mbox{max}[(\rho/\sigma_{\rho}); (t_{0},t_{\mathrm{dur}})],
\end{equation} 
where we calculate $\rho/\sigma_{\rho}$ for every template and all pixels, and then, for each pixel, save the maximum value and the template that provided this value. Using the ratio $\rho/\sigma_{\rho}$ ensures that all the pixels are drawn from the same standard normal distribution. Note that $\rhonorm$ is the maximum of a Gaussian random variable, so its distribution is no longer Gaussian. The trials factor arising from the bank of templates is included in the $\rhonorm$ distribution through the maximization procedure, unlike the $\rho$ distribution for a single template where there is no need to correct for the trials factor. We handle the non-Gaussianity of the $\rhonorm$ distribution in Section~\ref{sec:performance}.

There are two more steps before transient identification: characterize the detection significance (reliability) and measure the detection efficiency (completeness). As we describe these steps in detail in Section~\ref{sec:performance}, we summarize them here. To characterize the detection significance, one designates a small part of the image as the playground region where there are assumed to be no transients, corrects the cumulative distribution of $\rhonorm$ in the playground region by the trials factor, i.e. the number of synthesized beams in the search region compared to the number in the playground region, and extrapolates the tail of this corrected distribution to a tolerable probability of false alarm. The extrapolated distribution determines the significance of any detection during the actual search. To measure the efficiency, one applies the detection threshold to the injected transients and computes the fraction that is recovered. This procedure handles non-Gaussianity in the data, such as artifacts or sidelobe confusion noise as well as the effects of maximization, thus making it a powerful technique. The pipeline implementation is described in Appendix~\ref{app:simetra}.


\section{Data Reduction}\label{sec:data}

The MWA is a low-frequency radio interferometer consisting of 128 aperture arrays known as ``tiles'' located at the Murchison Radio-astronomy Observatory in the Murchison Shire of Western Australia; for a complete description of the instrument, see \citet{tingay2013}. The data for this paper were taken according to the commensal MWA observing proposals\footnote{http://mwatelescope.org/astronomers/} G0009 (``Epoch of Reionisation,'' EOR) and G0005 (``Search for Variable and Transient Sources in the EOR Fields with the MWA'') for Semester 2013-B.  These observations were done using the ``point-and-drift'' strategy, where the primary beam pointing (beamformer setting) changed every $20$--$30$ minutes to track the field after it drifted across the field of view of the instrument. 

We used 1251 snapshot observations of the EOR0 field, which was centered on $(\mathrm{RA}, \mathrm{Dec}) = (0\degr, -27\degr)$, taken on 18 nights between 2013 September~2 and November~30. We included only snapshots taken when the field center was $<20\degr$ from the zenith, corresponding to $\sim 2$\,h of observation each night, and excluded data from 2013~October~15 because of known ionospheric activity \citep{loi2015}. Each snapshot is a multi-frequency synthesis image integrated over 112\,s with a bandwidth of 30.72\,MHz centered on 182.40\,MHz. Table~\ref{tb:obsummary} lists a summary of the observations. 

\begin{deluxetable}{ccc}
\tabletypesize{\footnotesize}
\tablecolumns{3}
\tablewidth{0pt}
\tablecaption{Summary of observations used in our analysis.}
\tablehead{
  \colhead{Date} &
  \colhead{Time Range (UT)} &
  \colhead{Number of Snapshots}}
\startdata
    2013-09-02 & 16:08:08--17:39:36 & 46 \\
    2013-09-04 & 16:00:16--18:32:48 & 74 \\
    2013-09-06 & 15:52:22--18:24:54 & 72 \\
    2013-09-09 & 15:40:39--18:13:03 & 75 \\
    2013-09-11 & 15:32:47--18:05:11 & 75 \\
    2013-09-13 & 15:24:55--17:57:19 & 72 \\
    2013-09-17 & 15:09:11--17:41:35 & 76 \\
    2013-09-19 & 15:01:19--17:31:43 & 71 \\
    2013-09-30 & 14:17:59--16:50:31 & 76 \\
    2013-10-02 & 15:07:03--16:42:39 & 47 \\
    2013-10-04 & 14:02:15--16:34:47 & 76 \\
    2013-10-08 & 13:46:31--16:19:03 & 76 \\
    2013-10-10 & 13:39:43--16:12:15 & 76 \\
    2013-10-23 & 12:48:39--15:21:03 & 76 \\
    2013-10-25 & 12:40:47--15:13:11 & 76 \\
    2013-10-29 & 12:25:03--14:57:27 & 75 \\
    2013-11-18 & 11:18:31--13:38:55 & 70 \\
    2013-11-29 & 11:30:41--12:56:09 & 42 \\
\enddata
\tablecomments{Each snapshot is centered on the EOR0 field, or $(\mathrm{RA}, \mathrm{Dec}) = (0\degr, -27\degr)$, and integrated over 112\,s at 182.4\,MHz.}\label{tb:obsummary}
\end{deluxetable}

\subsection{Preprocessing}

Raw interferometric data were converted into the \uvfits\ format \citep{greisen2012} by the \cotter\ MWA preprocessing pipeline \citep{cotter}. During this process, \cotter\ used \aoflagger\ (\citealt{offringa2010},\citealt{offringa2012}) to flag radio-frequency interference, frequency channels affected by bandpass aliasing (240\,kHz on each edge of a 1.28-MHz coarse channel), as well as known bad tiles, which might vary from night to night depending on the state of the instrument. To decrease the file size, \cotter\ also averaged the data to 1-s time resolution and 80-kHz frequency resolution. 

\subsection{Calibration}

We developed a data reduction pipeline based on the Common Astronomy Software Applications (CASA) package\footnote{http://casa.nrao.edu/} (v4.1.0) \citep{casa} and the widefield imager \wsclean\ \citep{wsclean}. 

We built a point-source sky model for each snapshot observation, using the 11 brightest point sources in the field after primary beam attenuation according to the MWA Commissioning Survey Catalog \citep{mwacs}. We generated the model visibilities and the calibration solutions in CASA, using in particular the tools \texttt{componentlist}, \texttt{ft}, \texttt{bandpass}, and \texttt{gencal}. Then we performed one iteration of phase and amplitude self-calibration with \wsclean. 

Finally, we averaged over many observations on the same night the calibration solutions generated in the previous step to produce a single calibration solution for this night. This was done in two steps: (1) we selected a list of observations for which the \aegean\ source finder (v951) \citep{hancock2012} detected at least 1500 sources in each 112-s snapshot as we found this to be a practical indicator of image quality; (2) we averaged the calibration amplitude and phase solutions for each tile, polarization, and frequency channel for the selected observations while ignoring the highest and lowest $10\%$ of the data.

This procedure provided stable and smooth calibration solutions, which were not expected a priori to vary significantly over time and frequency. It also gave more reliable estimates of the source flux densities. 

\subsection{Imaging}

\begin{figure}
  \centering
  \includegraphics[width=0.5\textwidth]{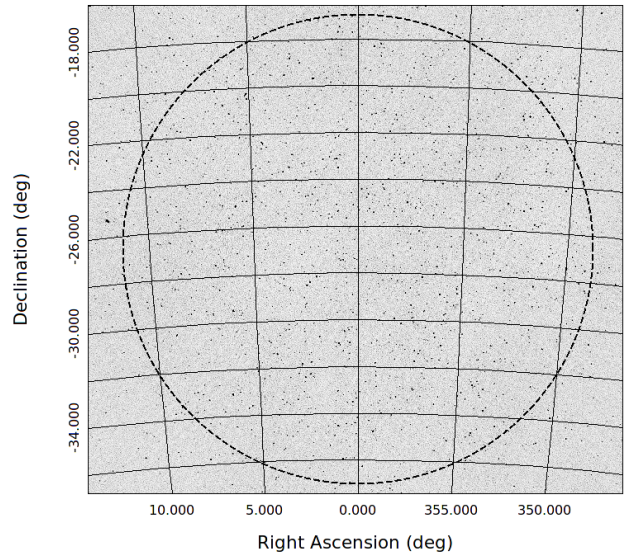}
  \caption[Example snapshot of EOR0]{An example snapshot of the EOR0 field. It was integrated over 112\,s and cleaned in the XX polarization, plotted with the J2000 coordinate grid and a squared color scale. The dashed circle has a radius of $10^{\circ}$, which is approximately the outer boundary of the field included in our analysis.}
  \label{fig:sky}
\end{figure}

After we applied the average calibration solutions to each snapshot, we generated multi-frequency synthesis images over $30.72$\,MHz bandwidth in the instrumental XX and YY polarizations, setting \wsclean\ to use uniform weighting that gave a synthesized beam of $\sim 2'$, a pixel size of $0.5'$, and an image size of $4096 \times 4096$ pixels, which corresponded to a field of view of $34\degr \times 34\degr$. Figure~\ref{fig:sky} shows an example snapshot of the EOR0 field. 

\subsection{Primary Beam Correction}

We adopted an empirical approach for primary beam correction. While there are theoretical primary beam models for the MWA \citep{sutinjo2015}, they are less accurate at frequencies $\gtrsim 180$\,MHz. We found that the source light curves extracted from images corrected by the theoretical primary beam model at $182$\,MHz showed systematic trends of $\sim 4\%$ flux change per hour that depended on the right ascension of the source. To alleviate this effect, we measured the empirical primary beam value at each source location by comparing the observed flux of the source to its catalog flux, which we assumed to be true and static, and fitted a smoothing spline to the measured beam values. This procedure reduced the systematic trends of the light curves to $\lesssim 2\%$ flux change per hour and removed most of the systematic variations in the light curves. A similar approach to empirical primary beam correction was done by \citet{thyagarajan2011} for the Very Large Array; for a more detailed discussion on the empirical primary beam we used, see Appendix~\ref{app:pbcor}. 


\section{Demonstration}\label{sec:performance}

In this section we demonstrate how a transient search with our matched filter technique proceeds and that it performs well on real data. Before we identify possible transient signals according to the matched filter detection statistic, we need to characterize the background distribution of $\rhonorm \equiv \mbox{max}(\rho / \sigma_{\rho})$. Using $\rhonorm$ instead of $\rho$ ensures that all pixels are drawn from the same distribution. Characterizing the background determines the significance of any detection and provides a threshold $\rhonorm^{\ast}$ for us to compare the performance of different searches and to measure the search efficiency. 

We used a small area ($\sim 10\%$) labeled as the ``playground region'' in the images for the background characterization. In this region, we assumed that there were no transient events; existing limits on the rate of radio transients at low frequencies suggest that these events are relatively rare (e.g. \citealt{jaeger2012},\citealt{rowlinson2015}), so our assumption should be valid. Significant transients would appear as a tail in the $\rhonorm$ distribution that we would examine further. 

\begin{figure}
  \centering
  \includegraphics[width=0.5\textwidth]{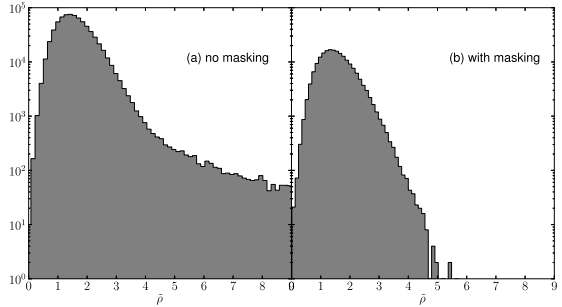}
  \caption[Distribution of $\rhonorm$ for the playground region]{(a) Distribution of $\rhonorm$ for all the pixels in the playground region, with the $x$-axis limited to $0 \leq \rhonorm \leq 9$. (b) The same distribution of $\rhonorm$ but without the ``bad'' pixels, i.e. the pixels in the region where the primary beam is poorly modeled and the pixels that contain bright sources affected by the primary beam systematics. The tail in the background distribution, which extends to $\rhonorm \sim 140$ in the left panel, is removed by masking those bad pixels.}
  \label{fig:rho_mask}
\end{figure}

To demonstrate our transient detection technique, we present the results for one light curve template: the top-hat with a duration of 15\,days and a brightness of $1$\,Jy\,beam$^{-1}$, searched over different start times $t_{0}$. In principle, we could have treated the observation time of every 112-s snapshot as a unique $t_{0}$, but since computation time scaled with the number of search parameters, we shifted $t_{0}$ by $\sim 10\%$ of the duration where there were data, which in this case corresponded to the start of every night of the observation. The distribution of $\rhonorm$ for the entire playground region is presented in Figure~\ref{fig:rho_mask}. 

As evident in Figure~\ref{fig:rho_mask}a, there is a significant tail in the $\rhonorm$ distribution. This is due to residual primary beam effects as the pixels in the tail are located in the regions where the primary beam is poorly modeled or that contain bright sources ($>100$\,mJy). We masked the pixels beyond $\sim 10^{\circ}$ from the phase center and excluded the source pixels by creating a $20 \times 20$-pixel square mask centered on the (RA, Dec) of each source. After masking, we managed to remove the tail in the distribution as illustrated in Figure~\ref{fig:rho_mask}b. The distribution after masking is what we work with and refer to as the ``background distribution.'' 

\begin{figure}
  \centering
  \includegraphics[width=0.5\textwidth]{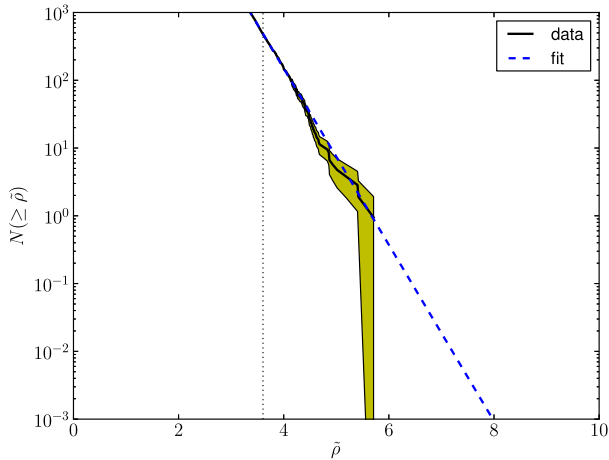}
  \caption[Cumulative background distribution of $\rhonorm$]{Cumulative background distribution of $\rhonorm$ (black solid line). The shaded area is the error region as computed from $\sqrt{N}$. Only the tail of the distribution (500 data points to the right of the vertical dotted line) was used for fitting. The fit (blue dashed line) is an exponential function: $f_{N}(\rhonorm) = \hat{N}\exp(-\rhonorm/\hat{\rho}) = 2.38 \times 10^{7} \exp(-\rhonorm/0.33)$. The reduced $\chi^{2}$ for the fit is 0.25. The fit was then used to compute the false alarm probability $\pfa$; see text for the details of the calculation. At $\pfa = 10^{-3}$, the transient detection threshold is $\rhostar = 7.98$.}
  \label{fig:rho_cdf}
\end{figure}

The background distribution of $\rhonorm$ lets us derive the probability of false alarm $\pfa$ (equivalent to reliability) for different values of $\rhonorm$. Here $\pfa$ is the probability that our experiment contains a false positive. Every experiment has its own observation timescale and sky coverage, which need to be factored accordingly. We start with a quantity closely related to $\pfa$: the number of background detections $N(\ge \rhonorm)$ above a particular threshold. We scale $N(\ge \rhonorm)$ as derived from the playground region to the search region according to the number of synthesized beams for the different sky areas, thus accounting for the trials factor correctly. The trials factor from the number of templates is already accounted for in the background distribution when we maximized over different templates. Then we fit a smooth function to the tail of $N(\ge \rhonorm)$ to extrapolate the background rate for larger values of $\rhonorm$ where we might not have measurements as the playground region contains much less data than the search region. Non-Gaussianity and non-thermal components of the image noise, such as sidelobe confusion and image artifacts, are modeled in this empirical fit. We determine the fit parameters by minimizing the negative log-likelihood for a Poisson distribution, where $N(\ge \rhonorm)$ is treated as the Poisson mean. As illustrated in Figure~\ref{fig:rho_cdf}, an exponential function of the form $f_{N}(\rhonorm) = \hat{N}\exp(-\rhonorm/\hat{\rho})$, where $\hat{N}$ and $\hat{\rho}$ are the fit parameters, fits the tail of $N(\ge \rhonorm)$ well within the errors $\sigma_{N} \equiv \sqrt{N}$. The probability for $N(\ge \rhonorm)$ to be non-zero, assuming a Poisson distribution, is $P(N > 0) = 1 - P(N = 0) = 1 - e^{-N}$. We take $\pfa = P(N > 0)$, and since we require $\pfa \ll 1$, $\pfa = P(N > 0) \approx N(\ge \rhonorm)$. Having determined $f_{N}(\rhonorm)$, we choose a tolerable value of $\pfa$ (for example $\pfa = 10^{-3}$) and solve for $\rhostar$ such that $\pfa = f_{N}(\rhostar) = N(\ge \rhostar)$. For our exponential function, the threshold is of the form
\begin{equation}\label{eq:threshold_form}
\rhostar = \hat{\rho}\ (\log \hat{N} - \log \pfa).
\end{equation}

The threshold $\rhostar$ depends on how much of the distribution tail is used in the fit. For Figure~\ref{fig:rho_cdf}, we fitted the tail 500 points, which gave $\rhostar = 7.98$ at $\pfa = 10^{-3}$. When we fitted 100 points, we obtained $\rhostar = 7.37$, whereas when we fitted 50 points, we obtained $\rhostar = 7.26$. This dependence implies that a single threshold value is not very reliable for identifying transient events near the threshold, so in the actual search, we opted for the ``loudest event statistic'' (\citealt{brady2004},\citealt{biswas2009}), which we describe in more detail in Section~\ref{sec:analysis}. Nevertheless, the threshold is useful for comparing how well different searches perform, for providing an estimate of the brightness sensitivity, and for verifying the recovery of injected transients.

\begin{figure}
  \centering
  \includegraphics[width=0.5\textwidth]{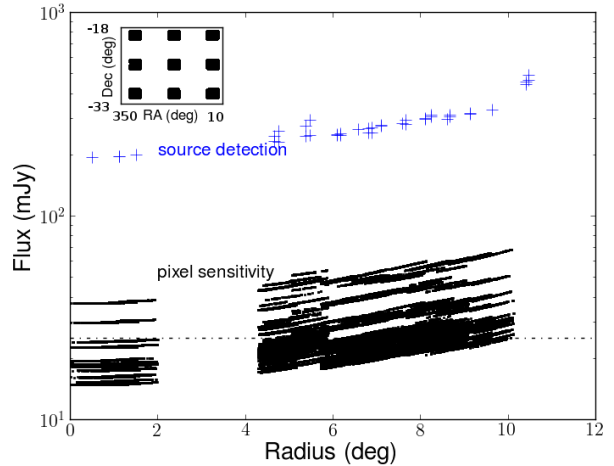}
  \caption[Flux density sensitivity comparison]{Flux density sensitivity comparison between the matched filter transient detection technique (using a 15-day top-hat template) and a source detection technique for the same set of 2-min snapshots. Each black dot is $A^{\ast} = (\rhostar / \sigma_{\rho}) = (7.98/ \sigma_{\rho})$ of a pixel and each blue cross is $6\imnoise$ (an example source detection threshold) at a particular radius from the EOR0 field center. The dash-dot line is the median flux density sensitivity $25.0$\,mJy quoted in the text for the matched filter technique. The matched filter technique achieves a better sensitivity than a source-finding algorithm for the same set of images without the need to produce a deeper integration image. The sensitivity decreases away from the phase center because of increased noise, but the significance remains the same. The gap between $2^{\circ}$ and $4^{\circ}$ is because of the discontinuous playground region sampling; see the inset where black rectangles mark the playground regions, each of which consists of several smaller $86 \times 86$-pixel patches. $\imnoise$ is calculated independently for each patch, so the stripes in $A^{\ast}$ correspond to the patches where the noise properties are different.}
  \label{fig:amp_radius}
\end{figure}

The flux density sensitivity of the search can be calculated according to $A^{\ast} = \rhostar / \sigma_{\rho}$ (see Section~\ref{sec:theory}; note $\rhostar$ already contains a factor of $\sigma_{\rho}$ unlike $\rho$). Strictly speaking, $A^{\ast}$ is unitless, so one needs to multiply by the template $\mathbf{f}$ to convert it into brightness units, but we drop $\mathbf{f}$ for a simpler notation as we choose $\mathbf{f}$ to have units of $1$\,mJy\,beam$^{-1}$. Because $\sigma_{\rho}$ depends on the primary beam, as presented in Equations~\ref{eq:c0} and \ref{eq:rhonorm}, pixels closer to the edge of the primary beam have different values of $\sigma_{\rho}$ compared to the pixels closer to the center of the primary beam. This implies that we have non-uniform brightness sensitivity across the image, where we are more sensitive to fainter transient sources toward the center of the primary beam. This is illustrated in Figure~\ref{fig:amp_radius}, where we also show an example detection threshold $6\imnoise$ for a source finder. 

To quantify the flux density sensitivity in a single number, we computed $A^{\ast}$ separately for each pixel, then we computed the median value of $A^{\ast}$ integrated over 1\,beam. Since the distribution of $\sigma_{\rho}$ and hence $A^{\ast}$ is skewed, the median is a better estimator of the flux density sensitivity than the mean. However, we stress that this value only provides an estimate of how well the search might perform and should not be interpreted as a strict threshold because a single value for the flux density sensitivity is a convenience and cannot capture the complexity of the data. 

It is $\rhonorm$ that matters in this technique. Since our technique uses $\rhonorm$ and not flux density as a metric to identify transient sources, it is capable of detecting sources fainter than the median flux density sensitivity at the \textit{same significance} (reliability) but a lower efficiency (completeness), making it potentially more powerful than the techniques that apply a more stringent flux density threshold. For our clean images and a 15-day top-hat template, $\pfa = 10^{-3}$ corresponds to $\rhostar = 7.98$ and a median flux density sensitivity of $25.0$\,mJy; note that fitting 50 points instead of 500 gives $\rhostar = 7.26$, which corresponds to a median flux density sensitivity of $22.7$\,mJy, a value not very different from $25.0$\,mJy. 


\section{Analysis}\label{sec:analysis}

We ran three separate blind transient searches on the MWA data. Each search used a different set of light curve templates and thus was sensitive to transients on a different timescale. 

\subsection{Defining the Searches}

The timescales spanned by the data ranged from 2~min to 3~months, but we did not have uniform sensitivity over all possible timescales. Some timescales could not be probed because of gaps in the observations, while other particular timescales suffered from systematic effects, such as those due to the periodic change in the primary beam pointing. 

\begin{figure}
  \centering
  \includegraphics[width=0.5\textwidth]{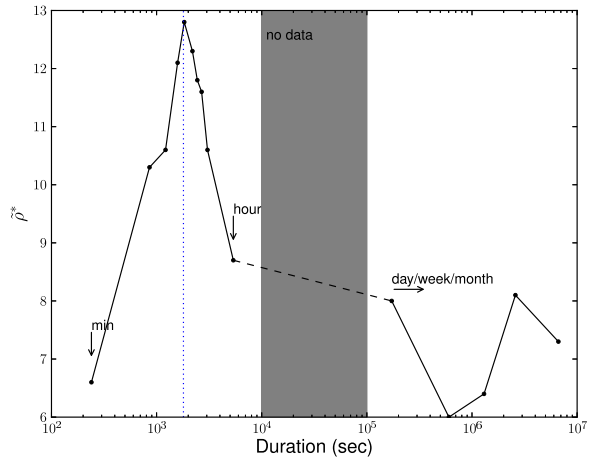}
  \caption[Expected thresholds for transients on different timescales]{Expected thresholds $\rhostar$ for transients on different timescales, characterized by the matched filter statistic, which depends on the image noise, the light curve template, and how the observations were taken. Smaller values of $\rhostar$ correspond to better sensitivities. The vertical blue dotted line marks the timescale for consecutive changes in the primary beam pointing; as we had poor sensitivity on this timescale, we excluded it from our search. The vertical grey region marks the approximate timescales that correspond to the gaps between observations where we had no data. Based on this plot, we divided our transient search into three parts: min, hour, and day-to-months.}
  \label{fig:search_definition}
\end{figure}

In order to determine which searches to run and which templates to use, we did a test run on the playground region to compare the expected transient thresholds for various timescales. Figure~\ref{fig:search_definition} shows the expected thresholds from the test run; we sampled roughly logarithmically between 2~min and 3~months but also sampled every $\sim 5$\,min between 15 and 50~min. There is a drastic increase in threshold around the 30-min timescale, which corresponds to the time between consecutive changes in the primary beam pointing. Hence we excluded that timescale from our search and defined the following three searches: minute, hour, and day-to-months. Despite the variation in $\rhostar$ values for the day-to-month templates, the median flux density sensitivities were about the same, so we grouped them together. The specific templates we used for the searches are listed in Table~\ref{tb:search}. Because of the primary beam systematics, we avoided templates with durations between $15$--$60$\,min. The $4$-min template is capable of recovering transients with durations $< 15$\,min as the difference between $\rho$ computed from the $4$-min template and $\rho$ computed from the perfectly matched template is $< 10\%$. As each night consisted of $\sim 2$\,hours of observation, we chose the $1.5$-hour template that is halfway between $1$ and $2$\,hours for the second search. Finally, for the last search, we considered all possible durations sampled by the data, excluding the gaps, and separated by at least $1$-day.

\begin{deluxetable}{llll}
\tabletypesize{\footnotesize}
\tablecolumns{4}
\tablewidth{0pt}
\tablecaption{Summary of our searches.}
\tablehead{
  \colhead{Search} &
  \colhead{Template} &
  \colhead{Duration} &
  \colhead{$\rhostar$}}
\startdata
    1: minute & top-hat & 4\,m & 7.6 \\
    2: hour & top-hat & 1.5\,h & 7.9 \\
    3: day-to-month & top-hat & 2\,d, 4\,d, 7\,d, 9\,d & 8.0 \\
    & & 11\,d, 15\,d, 17\,d, 28\,d & \\
    & & 30\,d, 32\,d, 36\,d, 38\,d & \\
    & & 51\,d, 53\,d, 57\,d, 77\,d, 88\,d & \\
\enddata
\tablecomments{$\rhostar$ is the threshold with $\pfa = 10^{-3}$.}\label{tb:search}
\end{deluxetable}

Each search consisted of three parts: (1) characterizing the background to determine the significance of any detection, (2) measuring the efficiency at which the transient events were detected, and (3) identifying the transient candidates if there were any. 

\subsection{Characterizing the Background}\label{subsec:search_threshold}

We characterized the background distribution of $\rhonorm$ by running the pipeline on the playground region. This region was $\sim 10\%$ of the image, which we assumed contained no transients. First we divided the inner $\sim 13\degr$ of the image, or $3096 \times 3096$\,pixels, into $86 \times 86$-pixel squares. This was because the pipeline ran on one CPU core allocated 3\,GB of RAM at any given time and would encounter memory issues if it processed more than $\sim 100 \times 100$ pixels each with $1251$ brightness measurements. Because of primary beam systematics, we only searched for transients in the inner $\sim 10\degr$, or $2096 \times 2096$ pixels, of the image. Then we chose the playground region to be nine $172 \times 172$-pixel patches divided into rows of three across the image. This choice sampled uniformly across the image to capture any spatial noise variation. 

As mentioned in Section~\ref{sec:performance}, we masked certain pixels to remove the tail in the background $\rhonorm$ distribution. One can consider the source mask as an ``auxillary channel'' for the search, where events that occur within a certain area of a (bright) source are vetoed. Brighter sources have larger sidelobes and hence a larger ``veto'' area. To define the veto area and refine the background characterization, we performed an empirical fit to determine the size of the source mask as a function of source flux density. We picked 6 sources with flux densities above $5$\,Jy, measured their sidelobe contamination areas in the unmasked transient sky map for the hour search, and fitted a straight line through the two points with the steepest slope to obtain the most conservative relationship between the size of the source mask and the source flux density:
\begin{equation}\label{eq:maskbox_fit}
\Delta n_{\mathrm{pix}} = 6.6 S + 0.5,
\end{equation}
where $\Delta n_{\mathrm{pix}}$ is the number of pixels to mask on each side of the source and $S$ is the value of the source flux density. If the fit returned $\Delta n_{\mathrm{pix}} < 10$, however, we set $\Delta n_{\mathrm{pix}} = 10$ to give a $20 \times 20$-pixel mask region. This is a conservative value to account for the size of the synthesized beam. We also manually flagged 31 double sources that were misidentified as single sources by \aegean. We note that by masking known, bright sources $> 100$\,mJy, we were no longer able to search for light curve variations in these sources although it would not affect our search for fainter transients. This was not ideal for a transient search, but it was necessary given the primary beam systematics in our data; with improvement on primary beam modeling and calibration techniques, it is conceivable that one does not need to mask those pixels.

\begin{figure}
  \centering
  \includegraphics[width=0.5\textwidth]{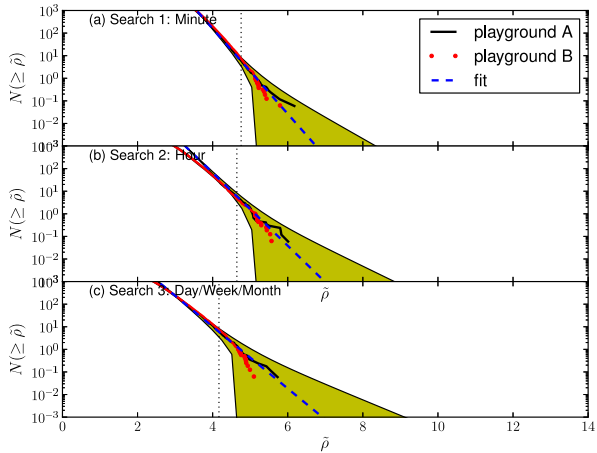}
  \caption[Background distribution of $\rhonorm$ for the playground region]{Cumulative background distribution of $\rhonorm$ for the playground region for the three searches. The blue dashed line is an exponential fit to the tail of the playground $A$ distribution, where the tail consists of 100 points to the right of the vertical dotted line. The yellow shaded region is the $\sqrt{N}$ error region for the fit, showing that the distributions from playground $A$ and $B$ agree within the error bars. All curves are normalized to the number of synthesized beams in playground $B$. The fit determines $\rhostar$ at a predetermined $P_{FA} = N(\ge \rhonorm)$ for $N(\ge \rhonorm) \ll 1$ when it is normalized to the number of synthesized beams in the search region. See text for further discussion.}
  \label{fig:playground_verification}
\end{figure}

We then divided the playground region into two parts, $A$ and $B$, by choosing alternating $86 \times 86$-pixel squares. Playground $A$ was used to characterize the background distribution of $\rhonorm$ and to set a threshold $\rhostar$ by extrapolating the tail of the distribution to our choice of false alarm probability $P_{FA}$ (the reliability of detected candidates), while playground $B$ was used to verify that extrapolation and the trials factor normalization. The extrapolation was done by fitting an exponential function to the tail of the cumulative distribution of $\rhonorm$ as described in Section~\ref{sec:performance}. For all three of our searches, we chose $P_{FA} = 10^{-3}$ for $\rhostar$, which meant that the probability of detecting a false positive in each search (experiment) is $\leq 10^{-3}$. If the distribution is Gaussian, which it is not, this probability corresponds to a significance of $3.3$-$\sigma$. Figure~\ref{fig:playground_verification} shows the extrapolation and verification for the three searches. While the normalization was set to the number of synthesized beams (independent pixels) in playground $B$ for the verification process, the threshold was determined after normalizing the background distribution to the number of synthesized beams in the search region. The values of $\rhostar$ for the three searches are listed in Table~\ref{tb:search}. 

We mention that another possible way to determine the background distribution is to shuffle the images in time and then use the entire image instead of defining a playground region. This avoids the need to extrapolate the tail of the $\rhonorm$ distribution, but we caution that random shuffling could break any temporal correlations in the systematic errors and change the noise distribution. We did not do this in our analysis. Future work is necessary to determine the timescale on which to shuffle the images that would preserve the true noise distribution. 

\subsection{Measuring the Efficiency}

The efficiency (completeness) of each search characterizes the fraction of real transients successfully recovered and plays a role in the upper limit calculation of the transient surface density. We determined the efficiency by running the same search on the injection region. 

The injection region consisted of $10^{4}$ random pixels sampled over the entire image. For each pixel, we simulated a transient light curve by drawing randomly from uniform distributions of brightness, start times, durations, and other parameters that define the transient shape. The values for these parameters were identified by their pixel number and recorded in an injection file independent of and unknown to the search pipeline. Rather than simulating measurement uncertainties, we corrupted these simulated transient light curves by injecting them into the actual data, i.e. added the brightness of the simulated transient to the observed light curve in the corresponding pixel at the corresponding times. This preserved the noise properties of the real data. These light curves, with injected transient signals, were then processed through the pipeline in the exact same manner as the transient search itself. In other words, the pipeline did not know whether or not the light curves it was processing contained injections or not. The injection run remained blind in this manner and would not have biased the results. The pipeline returned a list of transient candidates that passed the same criteria applied to the search region. We identified these candidates by their pixel number and compared the transient parameters as recovered by the pipeline, which had been corrupted by noise in the real data, to their ``true'' values stored in the injection file. We note that gaps in the observed data made comparisons of start times and durations tricky as an injected transient might have a ``true'' start time that occurred during a gap in the observation. We corrected for this by identifying the time of the closest observed snapshot before the ``true'' start (or end) time and compared the ``effective'' start times and durations since a real transient that starts before an observation is indistinguishable from a transient that starts at the beginning of an observation.

For all three searches, we injected transients with the top-hat profile, sampling uniformly in transient duration, start time, and brightness (amplitude). For the day-to-month search, we also injected transients with the fast-rise-exponential-decay (FRED) profile to mimic what real transients might look like, e.g. radio flares from X-ray binaries \citep{lo2014}, sampling uniformly in characteristic rise and decay times, start time, as well as peak flux. All injection parameters are listed in Table~\ref{tb:injection}.

\begin{deluxetable}{llll}
\tabletypesize{\footnotesize}
\tablecolumns{4}
\tablewidth{0pt}
\tablecaption{Summary of our injection runs.}
\tablehead{
  \colhead{Injection} &
  \colhead{Template} &
  \colhead{Duration} &
  \colhead{Peak Flux}}
\startdata
    1: minute & top-hat & 2--12\,m & $<1300$\,mJy \\
    2: hour & top-hat & 1--2\,h & $<450$\,mJy \\
    3: day-to-month & top-hat & 1--90\,d & $<150$\,mJy \\
    & FRED & $\tau_{1} = 1$--2\,d, $\tau_{2} = 30$--40\,d & $<160$\,mJy \\
\enddata
\tablecomments{We sampled uniformly in durations and peak fluxes as well as start times (not listed) that spanned the entire 3 months of observation. For the FRED profile, $\tau_{1}$ is the characteristic rise time, and $\tau_{2}$ is the characteristic decay time.}\label{tb:injection}
\end{deluxetable}

\begin{figure}
  \centering
  \includegraphics[width=0.5\textwidth]{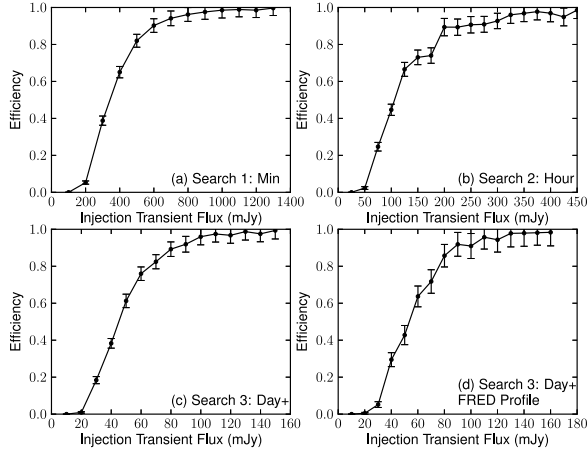}
  \caption[Efficiency for the three searches]{Efficiency (completeness) of recovered transients for the three searches. The efficiency increases at lower flux densities as the searched duration increases because longer durations imply longer integration times and lower image noise. Panel (d) shows the efficiency for when the transients were injected with the FRED profile, which is qualitatively different from the top-hat templates used in the search. It is not drastically different from the efficiency for when the transients were injected with the top-hat profile as in panel (c), demonstrating that the top-hat template is capable of recovering transients that are not top-hat in shape.}
  \label{fig:sr_efficiency}
\end{figure}

After running the search on the injection region, we applied a cut on $\rhonorm$, choosing only the events with $\rhonorm \ge \rhostar$. These were the recovered transient events. By computing the ratio of the number of recovered events to the total number of injected events, we determined the efficiency as a function of flux density for each search, as shown in Figure~\ref{fig:sr_efficiency}. Since sensitivity improves with lower image noise or longer integration time, the day-to-month search was able to recover fainter transient sources at a higher efficiency than the minute or hour search. 

\begin{figure}
  \centering
  \includegraphics[width=0.5\textwidth]{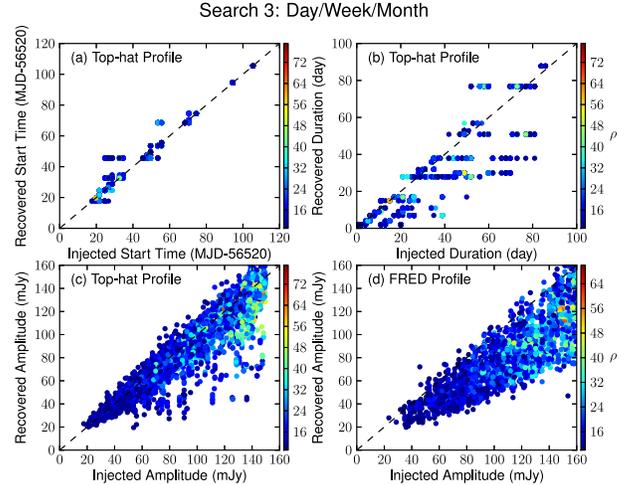}
  \caption[Recovery transient parameters]{Recovery of transient parameters. The pipeline is able to recover injected transient parameters fairly accurately; see also Figure~\ref{fig:param_accuracy}. There is a slightly bigger spread in the recovered durations, but that is because the search used a limited set of durations compared to the uniform sampling of injected durations. The flux densities recovered for the FRED profile injections are systematically lower than the injected peak fluxes, and that is a result of the qualitative difference between the injected and the searched light curve profiles; the FRED profile is sharper than the top-hat profile so the recovered flux density is smeared out. Although we only show this for the day-to-month search, the pipeline performs similarly for the other two searches.}
  \label{fig:recovered_parameters}
\end{figure}

\begin{figure}
  \centering
  \includegraphics[width=0.5\textwidth]{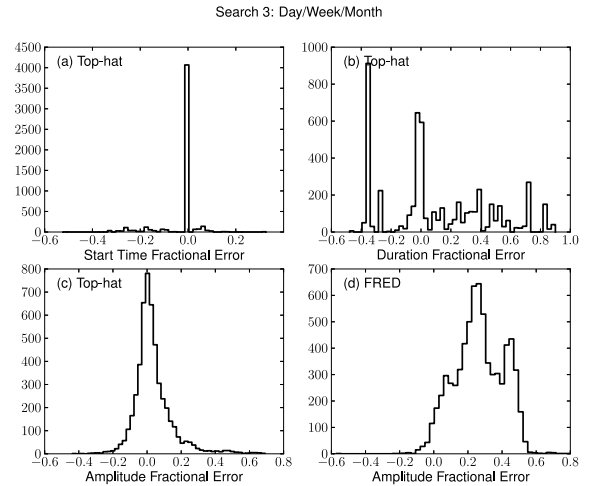}
  \caption[Accuracy of recovered transient parameters]{Accuracy of recovered transient parameters, corresponding to the panels in Figure~\ref{fig:recovered_parameters}. The accuracy of each light curve parameter $p$ is quantified by its fractional error: $(p_{\mathrm{inj}} - p_{\mathrm{rec}}) / p_{\mathrm{inj}}$ except for the start time, which is characterized by the start time difference relative to the injected duration: $(t_{0,\mathrm{inj}} - t_{0,\mathrm{rec}})/(t_{\mathrm{dur},\mathrm{inj}})$. The means of the distributions are (a) $-1.9\%$, (b) $10.2\%$, (c) $3.3\%$, (d) $26.3\%$; and the standard deviations are (a) $7.8\%$, (b) $35.5\%$, (c) $11.6\%$, (d) $14.3\%$.}
  \label{fig:param_accuracy}
\end{figure}

Since we also knew the true injection parameters, we checked the accuracy at which the pipeline recovered these parameters, as shown in Figures~\ref{fig:recovered_parameters} and \ref{fig:param_accuracy}. The pipeline was able to recover injected parameters fairly accurately ($\lesssim 10$--$30\%$ $1$-$\sigma$ errors), even when the search was run with top-hat templates on injected transients with the FRED profile. This demonstrates that we are capable of detecting real transients in the data, if they are present, despite using simple top-hat templates for the search. This is because a top-hat template will always have significant overlap with a transient signal that rises then falls on a similar timescale, regardless of the exact shape of the transient signal (see Equation~\ref{eq:rho}). However, the significance of a detection from a template that does not match exactly will be less than that from an exact match, as shown from the lowered efficiency and accuracy of recovering FRED profiles using top-hat templates. On the other hand, if there is a detection, one can rerun the search with better-matched templates to measure the transient properties more accurately. 

\subsection{Identifying the Candidates}

\begin{figure}
  \centering
  \includegraphics[width=0.5\textwidth]{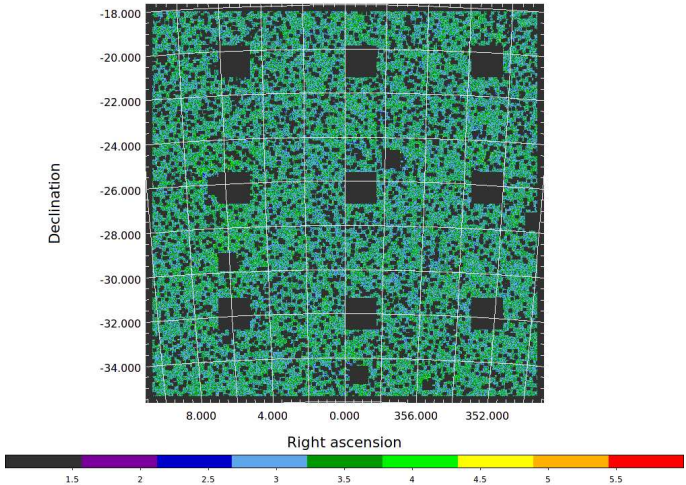}
  \caption[Map of the transient sky]{Example map of the transient sky, where the value of each pixel is $\rhonorm$ instead of brightness. This is for the minute search, so it shows the variability across the image on the minute timescale. The largest nine black patches make up the playground region and are not part of the map. The other smaller black squares are masked pixels, regions excluded from the search because they surround radio sources with flux densities $> 100$\,mJy.}
  \label{fig:rhosky}
\end{figure}

Finally, we ran the pipeline on the search region, which contained $\sim 3 \times 10^{6}$ image pixels or $\sim 2 \times 10^{5}$ synthesized beams, where one synthesized beam contained $\sim 16$ image pixels. We applied the cut $\rhonorm \ge \rhostar$ on individual pixels to identify transient candidates. If multiple pixels passed the cut, we grouped the adjacent ones together and considered them as one candidate because the image was oversampled. Since we computed $\rhonorm$ for every pixel, we could produce a ``transient'' sky map, as shown in Figure~\ref{fig:rhosky}, where each pixel contains the corresponding $\rhonorm$ instead of brightness. This map visualizes the variability on a particular timescale across the image, and a transient candidate would stand out as a source. 

\begin{figure}
  \centering
  \includegraphics[width=0.5\textwidth]{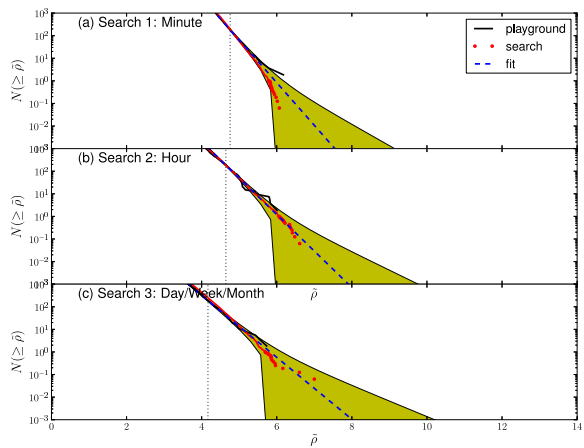}
  \caption[Distribution of $\rhonorm$ for the three search]{Cumulative distributions of $\rhonorm$ for the three searches. This is similar to Figure~\ref{fig:playground_verification}, but instead of comparing two playground regions, we compare the search and the playground regions. The tail to the right of the vertical dotted line was used for the fit. The searches are consistent with the background expectation and returned no transient candidates. See Table~\ref{tb:loudest} for a summary of the loudest (brightest) events.}
  \label{fig:search_result}
\end{figure}

We found no transient candidates. The properties of the loudest events are listed in Table~\ref{tb:loudest}. The $\rhonorm$ distributions from the search region agree very well with the background expectation as shown in Figure~\ref{fig:search_result}.

\begin{deluxetable*}{lllllll}
\tabletypesize{\footnotesize}
\tablecolumns{7}
\tablewidth{0pt}
\tablecaption{Summary of the loudest (brightest) events.}
\tablehead{
  \colhead{Search} &
  \colhead{$\rhonorm$} &
  \colhead{RA (deg)} &
  \colhead{Dec (deg)} &
  \colhead{Amp (mJy)} &
  \colhead{Duration} &
  \colhead{Start Time (MJD)}}
\startdata
    min & 6.1 & 355.16 & -18.34 & 220.8 & 4\,m & 56539.72 \\
    hour & 6.6 & 353.60 & -32.41 & 58.3 & 1.5\,h & 56614.52 \\ 
    month & 7.0 & 351.64 & -25.86 & 18.7 & 30\,d & 56537.67 \\
\enddata \label{tb:loudest}
\end{deluxetable*}

\subsection{Limits}

As we did not detect any transient candidates, we placed an upper limit on the transient surface density. We based our upper limit calculation on the ``loudest event statistic'' as derived by \citet{brady2004} and \citet{biswas2009}, which meant that we used the largest observed $\rhonorm$ instead of the search threshold $\rhostar$ to determine our search efficiency. This formalism does not rely on the threshold or the extrapolation described in Section~\ref{sec:performance} but on the brightest event actually observed in the data, making it independent of a threshold that could be set differently, and is very similar to the two-epoch equivalent snapshot rate introduced by \cite{bower2007}, except that it takes into account the search efficiency. We note that our upper limit formulation is only applicable to the case of no detections since we detected no transient events, although the loudest event statistic in its full formulation can be used in the case of a detection (see \citealt{brady2004} and \citealt{biswas2009}).

The probability that we detect no events above $\rhonorm$, assuming that the number of astrophysical transient events with a certain flux density is described by a Poisson distribution, is 
\begin{equation}
P(\rhonorm) = e^{-\mu \epsilon(\rhonorm)},
\end{equation}
where $\mu = \Sigma \Omega N_{e}$ is the Poisson mean, $\Sigma$ is the transient surface density, $\Omega$ is the area of each searched image, $N_{e}$ is the number of epochs or independent time samples, $\epsilon(\rhonorm)$ is the search efficiency as a function of flux density evaluated at $\rhonorm$. The upper limit on $\Sigma$ at a particular confidence level $p$ is then determined by $P(\tilde{\rho}_m) = 1-p$ or
\begin{equation}
\Sigma_{p} = - \frac{\ln(1-p)}{\Omega N_{e} \epsilon(\tilde{\rho}_m)},
\end{equation}
where $\tilde{\rho}_m = \mathrm{max}(\rhonorm)$ is the loudest event statistic. 

We computed the upper limit at $95\%$ confidence level separately for each of our three searches as they probed different timescales that corresponded to different astrophysical sources or processes. We determined $\Omega$ by multiplying the number of pixels searched and the area of each pixel, which gave us $\Omega = 186$\,deg$^{2}$. We determined $N_{e}$ by dividing the whole observation period, i.e. the time between the first and the last snapshot, by the transient duration, or the timescale of the search; if there were gaps in the observation that were as long as the transient duration, we subtracted the number of gaps from $N_{e}$. For our three searches, we have 625, 28, and 3 epochs respectively. 

\begin{figure}
  \centering
  \includegraphics[width=0.5\textwidth]{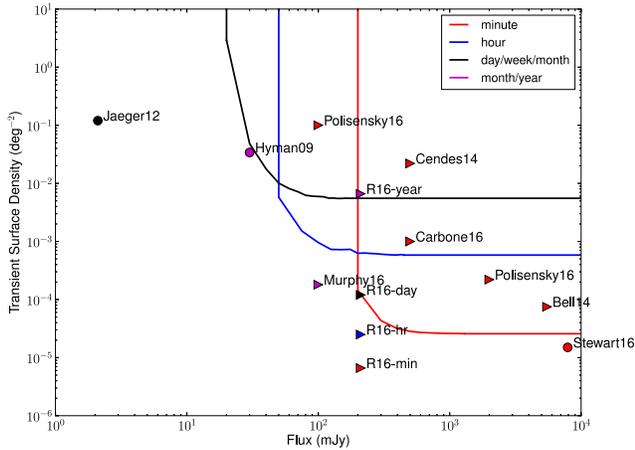}
  \caption[Upper limits on the transient surface density]{Upper limits on the transient surface density at $95\%$ confidence level from our analysis (solid lines) and other published results (symbols) between $150$ and $340$\,MHz. Note that each symbol should be interpreted as the vertex of an L-shaped upper limit line, which defines an exclusion region to the upper right. R16 denotes \citet{rowlinson2015}. Jaeger12 \citep{jaeger2012}, Hyman09 \citep{hyman2009}, and Stewart16 \citep{stewart2016} reported detections; the rest reported upper limits (\citealt{bell2014},\citealt{carbone2014},\citealt{cendes2014},\citealt{rowlinson2015},\citealt{polisensky2016},\citealt{murphy2016}). We only plotted two points for Polisensky16: the limit at the lowest flux density and the best limit reported for 10-min transients. Timescale is color-coded. Note that the detection reported by \citet{stewart2016} is at 60\,MHz.}
  \label{fig:limits}
\end{figure}

Figure~\ref{fig:limits} shows our results and compares them to the published results on the transient surface density between $150$ and $330$\,MHz. Our technique allows us to explore a larger phase space more efficiently. Despite using images each with a 2-min integration time, we achieved sensitivities equivalent to longer integration times for the longer duration transient searches. Although our results do not set more stringent limits at the same flux densities compared to \citep{rowlinson2015}, also an MWA result, their analysis covered a bigger sky area (452\,deg$^{2}$) and a longer observation period ($\sim 80$\,hr integration time spanning 1\,yr). If we naively scaled $\Omega$ and $N_{e}$ to match theirs, our limits would be comparable, e.g. $< 5.6 \times 10^{-6}$\,deg$^{-2}$ (ours at $100\%$ efficiency) compared to $< 6.6 \times 10^{-6}$\,deg$^{-2}$ for 4-min transients; we gained in the sense that our technique allows us to probe different timescales without the need to produce deeper integration images to achieve better sensitivities. The limits on these timescales will improve simply by adding more data, pushing toward lower and lower transient surface densities at the same flux density sensitivities. Pushing toward fainter flux densities would require better calibration techniques or primary beam modeling to decrease the image noise. Even with our current data, we reported improved limits and the best to date at 182\,MHz for flux densities between $\sim 20$--$200$\,mJy for hour- and month-long transients.

Our limits are also consistent with the reported detections of radio transients. The transient reported by \citet{jaeger2012} was much fainter than the sensitivity we could achieve with our data even though it occurred on a timescale that we probed ($\sim$\,day); if we assume a typical spectral index of $-0.7$, the source they detected at 2.1\,mJy at 325\,MHz would be 3.2\,mJy at 182\,MHz, which is an order of magnitude fainter than our best flux density sensitivity at $\sim 20$\,mJy. While our limit for the day-to-month search appears to overlap with the transient detection reported by \citet{hyman2009}, our results are still consistent with a non-detection. Their transient lasted about $\sim 6$\,months, which is longer than the total observation time of our data. Furthermore, their transient was detected near the Galactic Center, where it is plausible that the transient population and hence transient rate might be different from the extragalactic transient population, which we observed. If the transient population is similar, however, with more data, we should also begin to detect such transients. Likewise, with more data, we should be able to detect the transient population reported by \citet{stewart2016}, although their transient search was conducted at 60\,MHz, a lower frequency that might be probing different physical processes. 


\section{Conclusion}\label{sec:conclusion}

We developed a transient detection technique based on matched filters to search for transients in the presence of classical confusion noise. It searches for the light curve template that best matches the brightness variation above a constant signal in an individual pixel. The criterion for identifying transient candidates is set by the transient detection statistic $\rho$, which follows a well-defined distribution characterized by $\sigma_{\rho}$. The empirical background distribution of $\rhonorm \equiv \mbox{max}(\rho/\sigma_{\rho})$ determines the probability of false alarm $\pfa$ (reliability), which characterizes the significance of any detection and can be used to establish a threshold $\rhostar$ to compare the performances of different searches and verify the recovery of injected transients. 

For every pixel, $\rhostar$ can be converted to a flux density sensitivity according to $A^{\ast} = \rhostar / \sigma_{\rho}$. As different pixels have different noise properties, $A^{\ast}$ varies across the image, but the significance of the detection remains the same. The median flux density sensitivity, as computed from all the pixels, provides an estimate of the flux density sensitivity of the search but is not a strict threshold below which nothing is detectable. The efficiency (completeness) is determined by recovering injected transients with light curve parameters drawn from known or expected astrophysical distributions. 

We applied our technique for the first time to real data and demonstrated that our technique performs well despite the presence of residual sidelobe confusion noise and calibration errors. We performed an example search, using the MWA, for transients that have light curves resembling top-hats with a duration of 15\,days. For this type of transients, our technique is capable of detecting transients with fluxes $\sim 25.0$\,mJy at $\pfa \leq 10^{-3}$ for the experiment. As our technique identifies transient candidates by applying a cut on $\rhonorm$ and not flux, it remains sensitive to fainter transient sources at the same significance level but a lower efficiency. This is in contrast to flux-limited source detection techniques.

The ability to detect fainter transients in the presence of classical confusion noise increases the transient parameter space that a particular instrument can explore. As calibration techniques and primary beam modeling improve, one can push the limits of an instrument even further to study astrophysical transient sources that might be missed by source-finding algorithms. Our technique is also applicable to non-confusion-limited instruments and provides a way to study fainter source variations. 

We used this technique to search for transients in 3 months of MWA data. We ran three separate blind searches, using top-hat templates, to probe transients on different timescales: minute, hour, and day-to-month. For each search, we first characterized the background distribution of $\rhonorm$, which allowed us to set the threshold $\rhostar$ above which events were considered to be transient candidates. This threshold corresponded to a false alarm probability of $10^{-3}$, i.e. the probability that a candidate is a false positive (reliability) in the entire search is $<10^{-3}$. Then we characterized the efficiency of each search, or completeness, by running transient injections. The injections also demonstrated that we were able to recover transient properties accurately, even if the light curve profile of the injected transient differed qualitatively from the light curve template used in the search. 

We found no transient candidates. Thus we set an upper limit on the transient surface density for each of our searches. We took into account the search efficiency in our upper limit calculation, thus we were able to push to fainter fluxes than would otherwise be available. We reported improved limits at fluxes between $\sim 20$--$200$\,mJy for hour- and month-long transients, the best to date at 182\,MHz. This is consistent with reported transient detections in the literature, and it will easily improve with more data. 


\section*{Acknowledgements} \label{sec:acknowledgements}

This scientific work makes use of the Murchison Radio-astronomy Observatory, operated by CSIRO. We acknowledge the Wajarri Yamatji people as the traditional owners of the Observatory site. Support for the MWA comes from the U.S. National Science Foundation (grants AST-0457585, AST-0821321, PHY-0835713, CAREER-0847753, and AST-0908884), the Australian Research Council (LIEF grants LE0775621 and LE0882938), the U.S. Air Force Office of Scientific Research (grant FA9550-0510247), and the Centre for All-sky Astrophysics (an Australian Research Council Centre of Excellence funded by grant CE110001020). Support is also provided by the Smithsonian Astrophysical Observatory, the MIT School of Science, the Raman Research Institute, the Australian National University, and the Victoria University of Wellington (via grant MED-E1799 from the New Zealand Ministry of Economic Development and an IBM Shared University Research Grant). The Australian Federal government provides additional support via the Commonwealth Scientific and Industrial Research Organisation (CSIRO), National Collaborative Research Infrastructure Strategy, Education Investment Fund, and the Australia India Strategic Research Fund, and Astronomy Australia Limited, under contract to Curtin University. We acknowledge the iVEC Petabyte Data Store, the Initiative in Innovative Computing and the CUDA Center for Excellence sponsored by NVIDIA at Harvard University, and the International Centre for Radio Astronomy Research (ICRAR), a Joint Venture of Curtin University and The University of Western Australia, funded by the Western Australian State government. 

This research made use of Astropy\footnote{http://www.astropy.org}, a community-developed core Python package for Astronomy \citep{astropy2013}.


\appendix

\section{Pipeline Implementation}\label{app:simetra}

\begin{figure}
  \centering
  \includegraphics[width=1\textwidth]{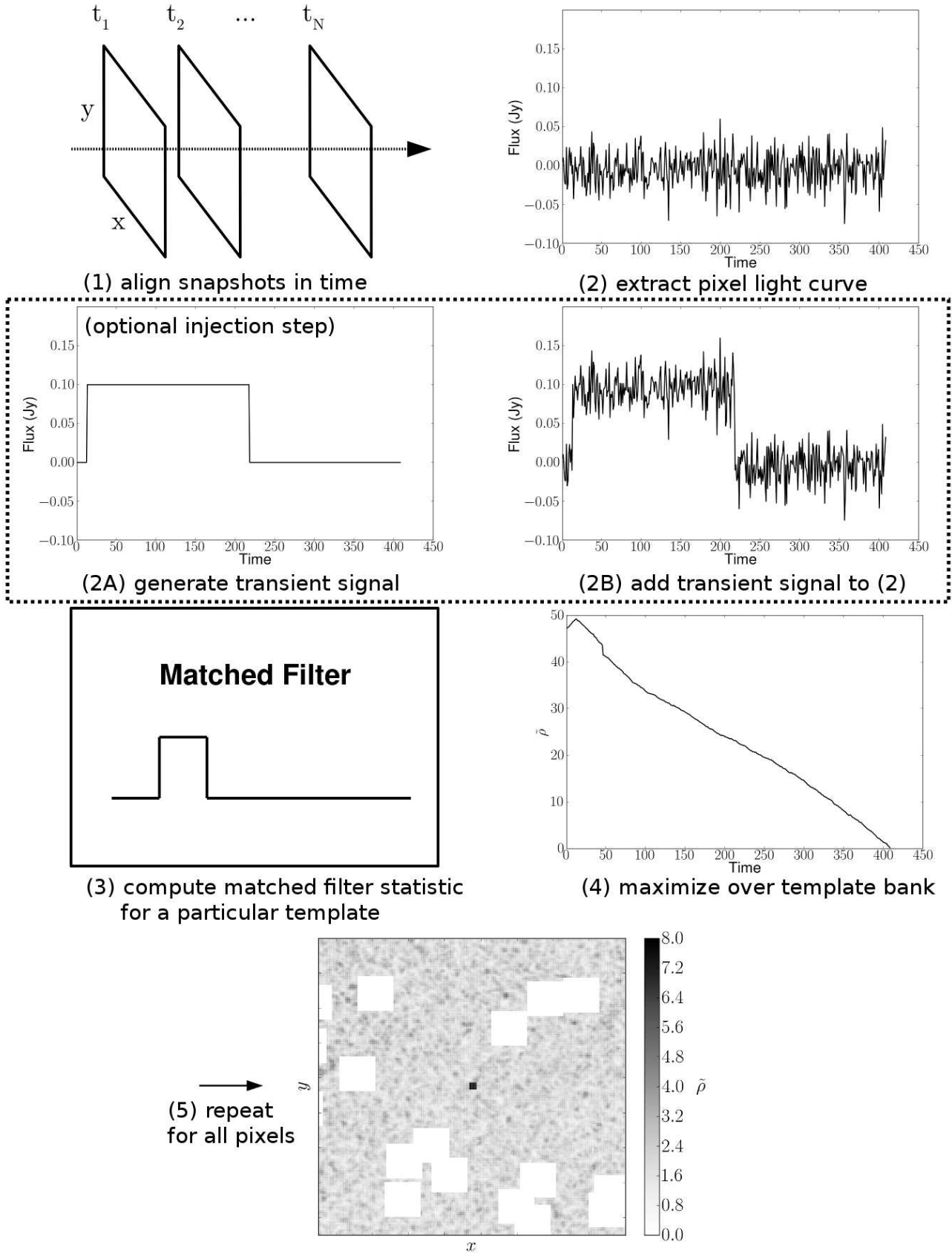}
  \caption[Illustration of \art]{Illustration of the \art\ pipeline. The color scale of $\rhonorm$ has been cut off at 8. The white boxes are the locations of masked bright sources. See text for details.
}
  \label{fig:pipeline_chart}
\end{figure}

We describe the \art\ pipeline, a \python\ implementation of the matched filter transient detection technique. The pipeline determines the transient detection statistic and the parameter values of the light curve template that best match the observed light curve for every image pixel. One can also use this pipeline to inject transient light curves with known parameters into the data before running the transient search; this determines the search efficiency. The pipeline is illustrated in Figure~\ref{fig:pipeline_chart}, and the code is available online\footnote{https://github.com/lufeng5001/simetra}. 

The pipeline takes, as input, a list of sky images, the corresponding primary beam images, and a choice of light curve template. At the moment of writing, two templates choices, the top-hat and the power-law, are available, but other choices are easy to implement as the code is designed to be as modular as possible. The pipeline reads the input \fits\ images and converts the flux density and primary beam values for every pixel into the time series $\mathbf{x}$ and $\mathbf{b}$, i.e. light curves for every pixel. Out of memory consideration, the pipeline only loads a subset of the image pixels each time. The pipeline then estimates the noise $\mathbf{\sigma}_{\thermal}$ in the sub-images by calculating the median absolute deviation and then converting that into the standard deviation to account for outliers in a robust manner. 

Here the user has the option to inject transients. The user chooses the type and the number of transient light curve templates to inject, and specifies the range and the distribution of the template parameters. Given these parameters, the pipeline generates the light curves on the fly and injects them into the data before it runs the transient search. We decided to inject transients directly into the pixel light curves instead of the visibilities because of computation concerns.

The transient search is the matched filter calculation. First the pipeline generates a phase space of template parameters, which can be different from the injection parameters depending on the user's choice. Then it iterates over every set of parameters, including all possible transient start times, generates the corresponding light curve template $\mathbf{f}$, and calculates $\rho$ and $\sigma_{\rho}$ according to Equations~\ref{eq:rho} and \ref{eq:rhonorm}. Gaps in the data are handled properly by sampling $\mathbf{f}$ at the existing image time stamps, and do not pose a problem. Finally the pipeline outputs a \fits\ table that contains the most significant $(\rho, \sigma_{\rho})$ and the corresponding template parameters for every pixel. As the amplitude can be determined by $A_{1} = \rho / \sigma_{\rho}^{2}$, it is not stored in the output file. 

The pipeline is highly parallelizable and not limited by computation as the matched filter calculation for each pixel is independent. I/O, i.e. converting the image \fits\ files into light curves stored as \texttt{numpy} \texttt{npz} files, is the bottleneck, but it only needs to be done once. Afterwards, the maximization over different start times is the slowest step and scales with the number of time samples. For $\mathcal{O}(10^{3})$ time samples, the computation time per $86 \times 86$ pixels, which was the smallest image unit we processed, for the entire pipeline was $\mathcal{O}(\mbox{min})$. We ran this process on 1 CPU core with $3$\,GB of allocated RAM on a computing cluster that comprised 14 computers with 99\,GB of RAM and 24 CPU cores per machine, each core operating at a speed of 60 MFLOPS. Running the pipeline for the full search of this paper, for example, can be completed in $< 1$\,week with 36 CPU cores. 

\section{Empirical Primary Beam}\label{app:pbcor}

The primary beam model establishes the flux scale during calibration and after imaging, so there have been many efforts to measure and model the primary beam of the MWA. For example, there is a project to map the beam pattern with an octocopter and a transmitter, while another used ORBCOMM satellites to measure the beam pattern of an MWA tile at 137\,MHz \citep{neben2015}. However, extrapolating this result to other MWA frequencies is not straightforward, so we need to rely on antenna modeling. 

The best available model for the MWA primary beam is that developed by \citet{sutinjo2015}. It improved on the previous model, which treated each antenna element as a Hertzian dipole, by incorporating mutual coupling between the elements and using an average embedded element pattern. This decreased the amount of instrumental Stokes leakage that was more prominent at frequencies $\gtrsim 180$\,MHz. However, the model beam assumed that all tiles were identical and unchanging over time, whereas in reality the MWA site is not perfectly flat and different tiles have different malfunctioning dipoles or beamforming errors \citep{neben2016}. In fact, we found that the source light curves extracted from the images corrected by the model beam at $182$\,MHz showed systematic trends of $\sim 4\%$ flux change per hour that depended on the right ascension of the source. As a result, we decided to measure and model the primary beam empirically to remove these errors, which more severely affected the higher frequency observations ($\gtrsim 180$\,MHz) as the Hertzian dipole approximation is less accurate than it is at $\lesssim 150$\,MHz where the MWA is designed to operate best.

If we assume that we know the true flux densities of the sources, we can determine the empirical primary beam according to the following relationship:
\begin{equation}
  b_{\mathrm{emp}, \mathrm{pol}}(\alpha, \delta) = \frac{S_{\mathrm{meas}, \mathrm{pol}}(\alpha, \delta)}{S_{\mathrm{ref}, \mathrm{pol}}(\alpha, \delta)},
\end{equation}
where $b_{\mathrm{emp}, \mathrm{pol}}$ is the empirically measured primary beam for a particular instrumental polarization (XX or YY), $S_{\mathrm{meas}, \mathrm{pol}}$ is the measured XX or YY flux density of a source with equatorial coordinates $(\alpha, \delta)$ and $S_{\mathrm{ref}, \mathrm{pol}}$ is the reference catalog (``true'') flux density of the same source. Similar analyses were done for the Very Large Array \citep{thyagarajan2011}. 

To measure the empirical beam, we used a fixed subset of sources instead of the entire ensemble detected in individual XX and YY snapshots. We did this for two reasons: (1) we wanted to ensure that the sources we used to measure the empirical beam have reliable flux density measurements, and (2) we wanted to avoid overfitting when we fitted a smooth function to the measured data points. 

For self-consistency, we used the MWA Commissioning Survey Catalog \citep{mwacs} as the reference catalog, which we also used for calibration. This avoided issues that could arise if we used source catalogs from other instruments, which might have different angular resolutions, frequency bands, sky coverage, and so on. We assumed the sources to be unpolarized and used the same catalog flux densities for both X and Y polarizations. 

When we selected the subset of sources, we filtered out the sources close to the null of the primary beam ($>13\degr$ from the phase center, corresponding to $\lesssim 0.3$ of the primary beam gain). We also filtered out the sources that appeared to have unreliable flux density measurements, which we determined by comparing the catalog flux densities $>1$\,Jy to the mean flux densities that we measured: if the mean flux density that we measured was $3$-$\sigma$ away from the catalog flux density, we removed it from the final catalog that we used to measure the empirical beam. As we could not measure a mean flux density without applying the primary beam correction to the images, we did this step iteratively by first using the model beam and then refining it with the empirical beam once. After this process, the reference catalog contained $245$ sources. The agreement between the catalog flux densities and the measured flux densities ensured that the fitting procedure used reliable data. 

For the fitting procedure, we assumed that the primary beam was smooth and fitted a smoothing spline to the empirical beam measurements on a snapshot-by-snapshot basis. We used biquartic splines $(k_{x}, k_{y}) = (4, 4)$ as they provided a better fit (lower residuals) than the default bicubic splines, whereas biquintic splines did not improve the fit significantly.

Fitting a spline function $s(x_{i}, y_{i})$ to a set of data $z_{i}$ that have measurement errors involves a trade between the smoothness of the spline and the goodness of the fit. The algorithm used by the \texttt{SmoothBivariateSpline} routine in \scipy\footnote{http://www.scipy.org/}, which we used for this analysis, determines the smoothest spline given the constraint that the goodness-of-fit is less than the smoothing factor $S$:
\begin{equation}
\sum_{i=1}^{m} w_{i} \left[z_{i} - s(x_{i}, y_{i}) \right]^{2} \leq S,
\end{equation}
where $z_{i}$ is the empirical beam measurement for each source, $(x_{i}, y_{i})$ is the equatorial coordinate of the source, and $w_{i}$ is the weight of each measurement \citep{dierckx1981}. We chose $S$ to be the number of sources that entered the fit (245), which was the default value, as it gave a satisfactory fit. We set $w_{i}$ to be the catalog flux density for each source, because the flux density errors derived from \aegean\ appeared to be too large to provide reliable inverse variance weights and did, in fact, make the fit worse. The unreliable errors reported by \aegean\ are a known issue and has since been fixed.

We also verified that the fitting function was reliable. Instead of deriving an empirical primary beam based on flux density measurements in clean XX and YY images before primary beam correction, we derived an empirical ``correction factor'' for the Stokes I images corrected by the model beam, using the same fitting procedure. Both procedures gave the same results, thus demonstrating that the fitting function was reliable. The empirical fitted beam removed most of the systematic errors in the light curves. It reduced the light curve slopes from $\lesssim 4\%$ to $\lesssim 2\%$ flux density change per hour and recovers (by design) a more accurate source flux density.


\bibliographystyle{apj}
\bibliography{fvh2016}

\end{document}